\documentclass[12pt]{article}

\usepackage[letterpaper,margin=1in,bottom=1.6in]{geometry}
\usepackage{amsmath,amssymb,mathtools}
\usepackage{amsthm}
\usepackage{bm}
\usepackage{graphicx}
\usepackage{array}
\usepackage{booktabs}
\usepackage{microtype}
\usepackage{xcolor}
\usepackage{natbib}
\usepackage{tabularx}
\usepackage{xurl}
\usepackage{xr-hyper}

\usepackage{setspace}
\usepackage{lineno}
\usepackage[hidelinks]{hyperref}

\newcommand{\argmin}{\operatorname*{arg\,min}}

\newtheorem{proposition}{Proposition}
\newtheorem{theorem}{Theorem}

\newtheorem{assumption}{Assumption}

\title{\Large LLM Benchmarking via Representation Multi-task Learning}
\author{Yuqing Xie\\
London School of Economics and Political Science
\and
Yuxuan Xu\\
London School of Economics and Political Science
\and
Yang Feng\\
New York University
\and
Yunxiao Chen
\\
London School of Economics and Political Science }
\date{}

\begin{document}
\maketitle
 
\begin{abstract}

Quantifying and evaluating the capabilities of Large Language Models (LLMs) remains a fundamental challenge. In this paper, we consider LLM evaluation based on their performance across items in multiple benchmark domains (e.g., mathematical reasoning and coding) within a leaderboard framework. Our goal is to address two core questions: (1) How do we derive more accurate domain-specific scores by borrowing information across domains? and (2) How do we define and estimate an overall score that aggregates performance across multiple domains? To solve these problems, we propose a statistical framework based on representation Multi-task Learning (MTL) and an item response theory model. Specifically, we define overall and domain-specific LLM traits through an Item Response Theory (IRT) model, and propose an MTL approach to estimate these traits from item-level response data. We develop a computationally efficient estimator and establish its minimax optimality under certain asymptotic regimes. This framework provides a rigorous measurement foundation for systematic LLM evaluation. We conduct extensive simulations, demonstrating the superior performance of the proposed method over competing methods. We apply the proposed framework to MMLU response data from the Hugging Face Open LLM Leaderboard, covering 4,272 LLMs and 13,232 items across 56 subjects. The empirical analysis reveals substantial heterogeneity in domain size and difficulty, together with strong positive cross-domain dependence, highlighting the value of the proposed approach.  

\medskip

\noindent\textbf{Keywords:} AI measurement science, representation multi-task learning, generalized latent factor model, item response theory, measurement reliability 
\end{abstract}

\setstretch{1.70}

\clearpage
\section{Introduction}

Artificial Intelligence (AI) systems, in particular Large Language Models (LLMs), are becoming increasingly powerful and are being deployed in high-stakes domains. Central to tracking this progress are public leaderboards that synthesize model performance across multiple benchmark domains, each comprising a large corpus of test items.
While these leaderboards provide vital empirical signals regarding model capabilities, current evaluation practices suffer from significant methodological shortcomings. They typically rely on ad hoc total scores to summarize an LLM's performance in individual domains, then aggregate them into a global score across all domains. This naive aggregation strategy presents several limitations. First, calculating domain scores independently ignores the strong correlations inherent in multi-domain LLM performance, missing an opportunity to statistically borrow information across domains for more reliable estimation. Second, simply aggregating domain-specific scores into an overall score may not provide a good summary of LLMs' general intelligence, as it again ignores dependence among domain-specific scores. Third, relying exclusively on raw scores is problematic because it fails to account for item difficulty, measurement error, and varying test characteristics, limitations long recognized in psychometrics and contemporary AI measurement science. 

This paper aims to introduce a principled statistical framework to resolve these multi-domain measurement challenges. The contributions of our work are twofold. First, 
we formulate a multidimensional Item Response Theory (IRT) framework  explicitly designed for cross-domain item response data. Within this architecture, we formalize the latent trait of general intelligence alongside domain-specific competencies, moving beyond naive aggregation to properly capture the complex hierarchy of LLM capabilities. Second, we develop a novel estimation procedure rooted in Multi-Task Learning (MTL). This approach allows us to efficiently borrow statistical strength across distinct benchmark domains when estimating domain-specific scores, while rigorously accounting for cross-domain heterogeneity when computing the overall global score. Furthermore, we establish the theoretical upper and lower bounds for the proposed estimators, providing theoretical justifications and guarantees. Under certain asymptotic regimes, the two bounds match, and thus the proposed estimator is minimax optimal.

Our work broadly connects to three strands of literature: psychometric modeling, representation multi-task learning, and high-dimensional generalized latent variable models.

\begin{enumerate}
    \item \textbf{Psychometric Item Response Theory Modeling.} 
Item Response Theory (IRT) constitutes a foundational psychometric paradigm widely utilized in measurement problems in social and behavioral sciences, especially education and psychology \citep{chen2025item}. The intellectual roots of IRT date back to the factor model for general intelligence introduced by \cite{spearman1904general}, who sought to extract a common intelligence factor underlying diverse cognitive tests. 
The development of classical unidimensional IRT models was subsequently formalized by foundational works, including \cite{rasch1960probabilistic}, \cite{birnbaum1968some} and \cite{lord1968statistical}.
These models have been further extended to 
the Generalized Latent Factor Model (GLFM) framework \citep{bartholomew1980factor, bartholomew2011latent,skrondal2004generalized} to allow for different types of item responses, and to multidimensional IRT models \citep{reckase2009multidimensional} to jointly model multiple latent traits. In recent years, IRT models have gained significant traction in AI evaluation \citep[e.g.,][]{kipnis2025metabench,schilling2025lifting,truong2026item,cai2025latent}. However, to the best of our knowledge, the existing literature has yet to address our methodological questions regarding the scoring of LLMs in a multi-domain setting, and the current work fills this gap.  

\item \textbf{High-Dimensional Generalized Latent Factor Models.} Classical IRT models concern data from human testing, where the number of test takers is often much larger than the number of items.  Under this setting, these models treat the test-taker-specific parameters as random variables and further make distributional assumptions on them \citep[see, e.g., ][]{chen2025item}. However, LLM benchmark data tend to depart from the data regime of standard human testing. In LLM benchmarking, LLMs serve as test takers, whose number may be smaller than that of human test takers in educational testing or psychological measurement. In addition, capability distributions may be skewed, clustered, or multimodal, making standard normality assumptions in classical IRT models unsuitable. Furthermore, LLM benchmark data involve substantially more items than human testing, leading to a high-dimensional regime in which the marginal maximum-likelihood estimator for classical IRT models may be unsuitable. 
To address issues with classical IRT analysis, we extend the high-dimensional GLFM analysis framework to the current multi-domain setting for LLM benchmarking. As considered, for example, in \cite{chen2019joint}, \cite{chen2020structured}, \cite{liu2023generalized}, and \cite{chen2022determining}, high-dimensional GLFMs pertain to IRT models and their extensions when both the number of test takers and the number of items are large. In these models, the test-taker-specific parameters are treated as fixed parameters and estimated together with the item-specific parameters. Theoretical guarantees for parameter estimation, including minimax optimality results, are established under the double asymptotic regime in which the number of test takers and the number of items both grow to infinity. The current work extends existing high-dimensional GLFMs to a multi-domain setting. We adapt technical tools from the high-dimensional GLFM literature to ensure theoretical rigor, while contributing back to that literature by broadening its modeling scope and analytical framework. 

\item 
 \textbf{Representation Multi-task Learning.} 
Representation MTL focuses on jointly learning shared feature spaces or latent representations across multiple related tasks to improve overall estimation efficiency and generalization performance \citep{baxter2000model, ando2005framework, maurer2016benefit,  tripuraneni2020theory, tripuraneni2021revovling, du2021fewshot, thekumparampil2021efficient, tian2025learning}. Rather than learning task-specific parameters in isolation, this paradigm leverages structural commonalities, such as shared low-dimensional subspaces, common linear representations, or coupled latent factors, allowing information to be borrowed effectively across tasks. By constraining or regularizing representations across tasks, the methodology achieves enhanced statistical adaptivity and robustness, effectively mitigating data scarcity in individual domains.  LLM benchmarking can be viewed as an MTL problem, where each domain can be seen as a task. The proposed method can be viewed as an MTL method that leverages similarities in LLMs' capability traits across different domains.  MTL methods are typically designed for prediction and inference tasks. To our knowledge, the current work is the first to apply an MTL approach to measurement. 

\end{enumerate}
  
The proposed method is applied to a dataset based on the Massive Multitask Language Understanding (MMLU) benchmark, which contains the responses from 4,272 LLMs to 13,232 items 
across 56 subject domains. Due to the heterogeneity of the subject domains and the complex dependence in the LLMs' performance across domains, 
accurately measuring the LLMs' overall and domain-specific capabilities is a nontrivial task. The proposed framework yields more reliable overall and domain-specific scores for the LLMs. 

The rest of this paper is organized as follows. Section~\ref{sec:data} introduces the dataset and outlines the primary research questions. Section~\ref{sec:method} presents the proposed methodology, followed by its theoretical properties in Section~\ref{sec:theory}. Section~\ref{sec:simulation} evaluates the finite-sample performance of the method through a series of simulation studies. Section~\ref{sec:real} demonstrates its practical utility via an empirical application to MMLU benchmarking, and Section~\ref{sec:discussion} concludes with final remarks and directions for future work. The Supplementary Materials contain additional theoretical, simulation, and empirical results.

\section{MMLU Data and Research Questions} \label{sec:data}

\begin{figure}[htbp]
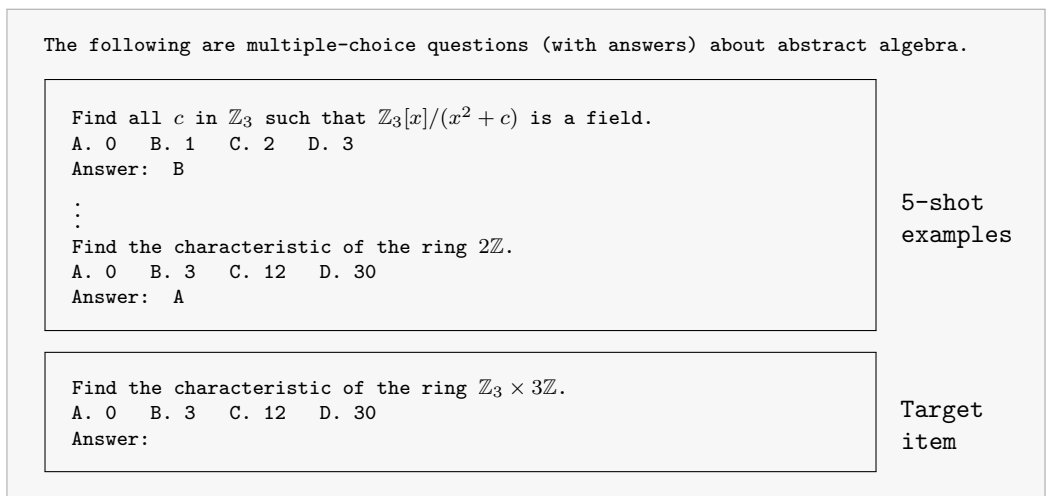

    \centering
    
    \setlength{\fboxsep}{10pt}
    \fcolorbox{black!35}{black!3}{
    \begin{minipage}{0.78\textwidth}
    \scriptsize
    \ttfamily
    The following are multiple-choice questions (with answers) about abstract algebra.\\
    
    \noindent
    \begin{minipage}[t]{0.84\linewidth}
    \fbox{\begin{minipage}{0.95\linewidth}
    Find all $c$ in $\mathbb{Z}_3$ such that $\mathbb{Z}_3[x]/(x^2 + c)$ is a field.\\
    A. 0 \quad B. 1 \quad C. 2 \quad D. 3\\
    Answer: B\\[0.45em]
    \vdots\\
    Find the characteristic of the ring $2\mathbb{Z}$.\\
    A. 0 \quad B. 3 \quad C. 12 \quad D. 30\\
    Answer: A
    \end{minipage}}
    \end{minipage}\hfill
    \begin{minipage}[t]{0.12\linewidth}
    \raggedright\footnotesize\textbf{5-shot examples}
    \end{minipage}
    
    \vspace{0.8em}
    
    \noindent
    \begin{minipage}[t]{0.84\linewidth}
    \fbox{\begin{minipage}{0.95\linewidth}
    Find the characteristic of the ring $\mathbb{Z}_3 \times 3\mathbb{Z}$.\\
    A. 0 \quad B. 3 \quad C. 12 \quad D. 30\\
    Answer:
    \end{minipage}}
    \end{minipage}\hfill
    \begin{minipage}[t]{0.12\linewidth}
    \raggedright\footnotesize\textbf{Target item}
    \end{minipage}

    \end{minipage}}
    \caption{An example of an MMLU five-shot prompt using abstract algebra items. The final item is the target question to be answered by the LLM.}
    \label{fig:mmlu-example-item}
\end{figure}

The proposed framework is motivated by the Massive Multitask Language Understanding benchmark \citep[MMLU;][]{hendrycks2021mmlu} for LLM evaluation. MMLU is a standard multiple-choice benchmark that evaluates LLMs across 57 diverse subjects, such as math, history, law, and computer science. In the current analysis, we excluded the miscellaneous subject, whose items are too heterogeneous in content to constitute a coherent domain, leaving 56 subjects. Our analysis is based on item-level response data from an archived version of the Hugging Face Open LLM Leaderboard\footnote{The item-level leaderboard data come from the archived Hugging Face dataset \url{https://huggingface.co/datasets/open-llm-leaderboard-old}}, where the item responses are dichotomous, i.e., whether an LLM answers an item correctly. After data preprocessing, the final analysis is based on 4,272 LLMs and 13,232 items from 56 subjects. Additional descriptive details on the evaluated LLMs are provided in Supplementary Materials C.1. On the one hand, the evaluated LLMs exhibit limited diversity in their provenance and underlying architectures. Only about 2\% of the models are published under the official accounts of major open-weight developers, including Meta, Mistral AI, Google, Alibaba, Microsoft, 01.AI, DeepSeek, TII, Cohere, AllenAI, Databricks, and Stability AI. The remaining 98\% consist of community-developed variants, such as instruction-tuned, preference-tuned, domain-adapted, and merged models. Thus, although the dataset contains many LLMs, the diversity of underlying architectures remains fairly limited, as most models are derived from a relatively small set of closely related backbones. Conversely, the sample does exhibit substantial variation in model scale. Among the models with detailed records in the leaderboard (3,966 out of 4,272 models), model scale ranges from 1B to 238B parameters, with a median parameter count of 7B. Models in the 7B range constitute the largest share of this subset, accounting for one half of the evaluated systems.
During the evaluation, each MMLU item is presented to the LLM in a five-shot, four-option multiple-choice format; an example prompt is shown in Figure~\ref{fig:mmlu-example-item}.

Two questions naturally arise regarding leaderboard-based benchmarking of LLMs using item-level performance data. First, how can we establish a standardized score for LLMs within each specific domain? Second, how do we derive an aggregate score that summarizes performance across all domains? At first glance, these appear straightforward and can be addressed by simple methods. However, due to several distinct patterns in the data, as discussed below, standard approaches fall short.

\begin{figure} 
    \centering
    \includegraphics[width=0.9\textwidth]{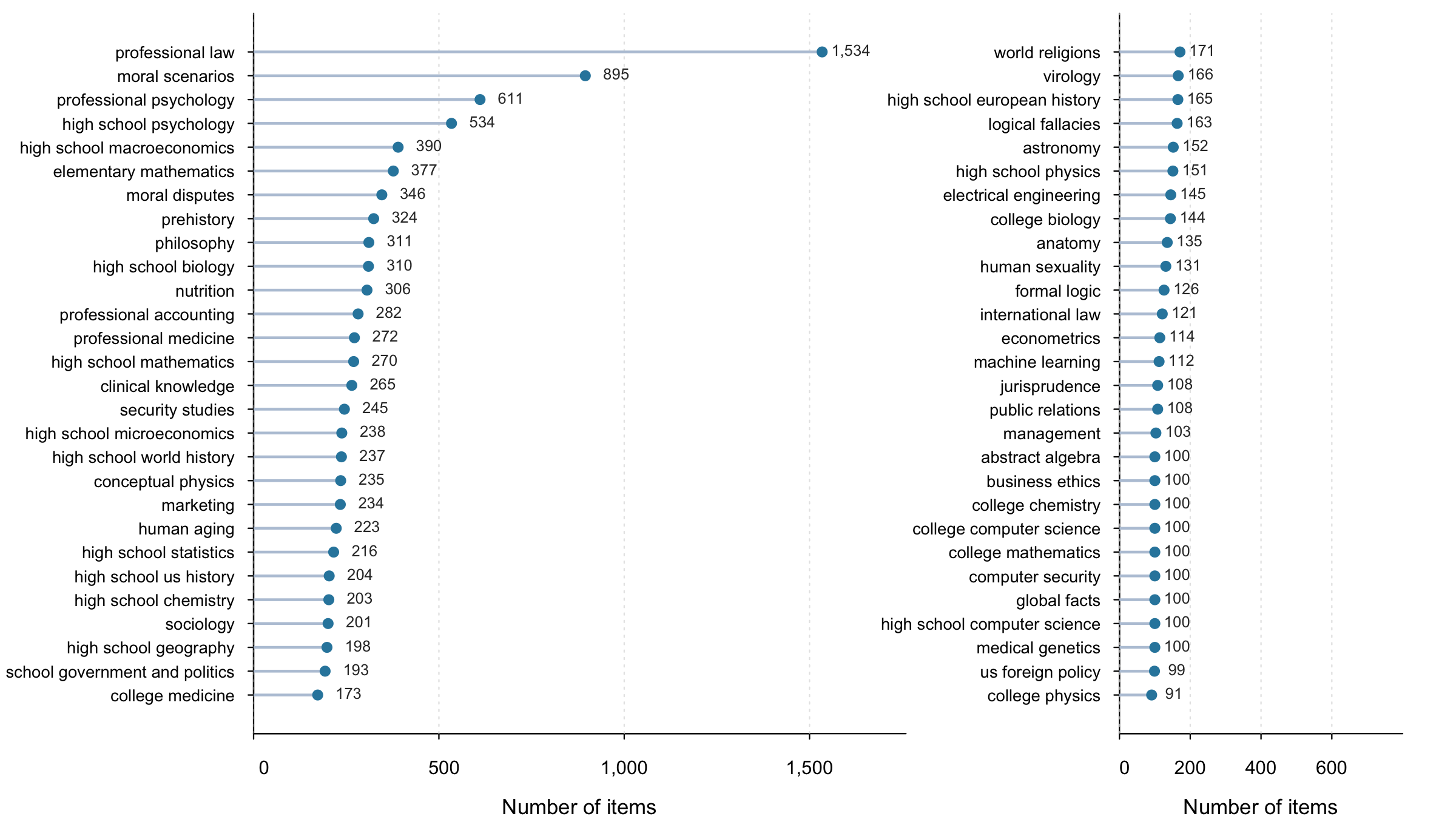}
    \caption{{The number of items in each of the 56 MMLU subjects, ordered by item count. For readability, the subjects are split across two panels; both panels display the same quantity and together cover all 56 subjects.}}
    \label{fig:mmlu-subject-item-counts}
\end{figure}

\begin{figure} 
    \centering
    \includegraphics[width=0.8\textwidth]{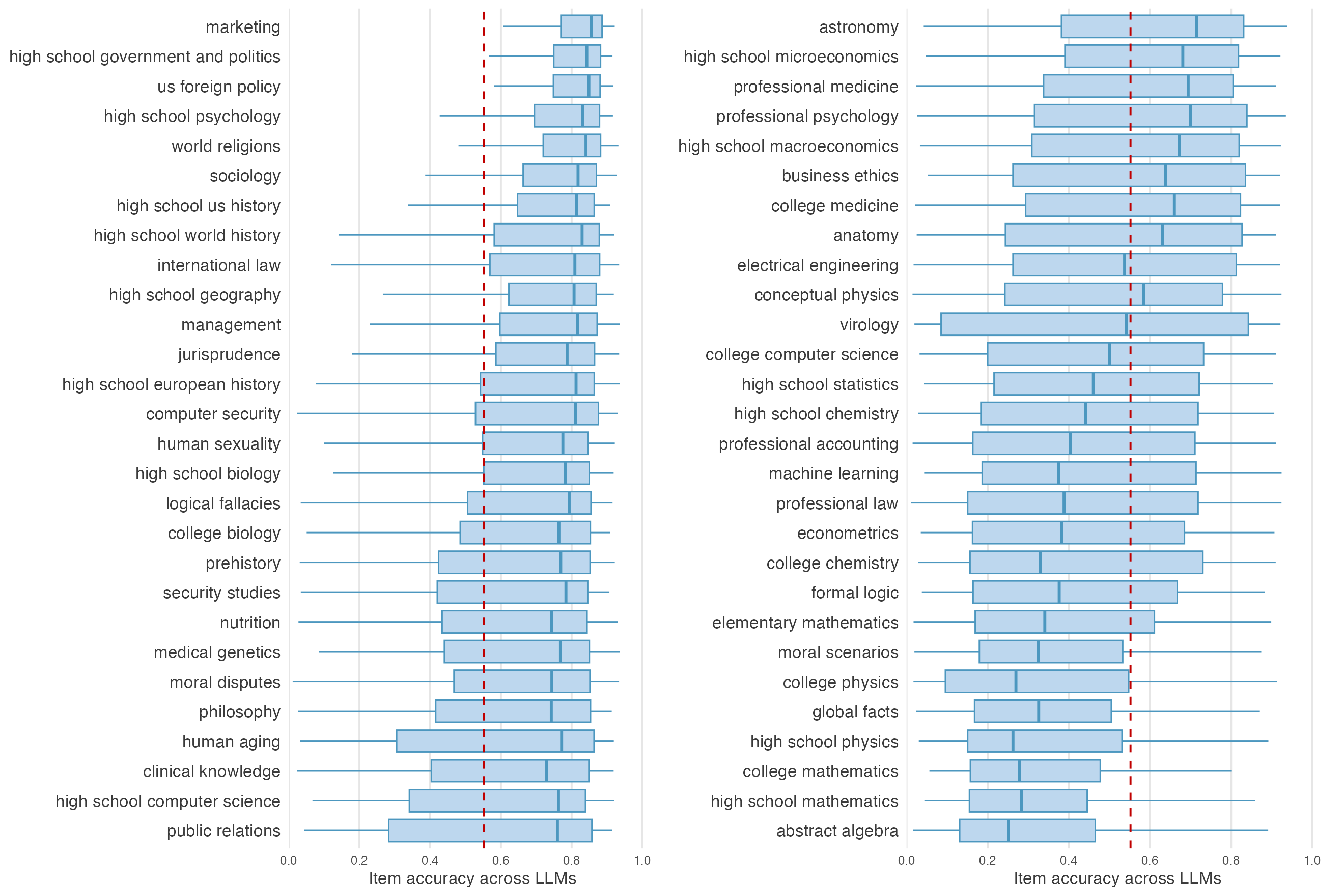}
    \caption{Distribution of item-level model accuracies within each MMLU subject. For each item, accuracy is the proportion of the 4,272 models that answer the item correctly.}
    \label{fig:mmlu-item-accuracy-by-subject}
\end{figure}

\begin{figure}[t] 
    \centering
    \includegraphics[width=0.8\textwidth]{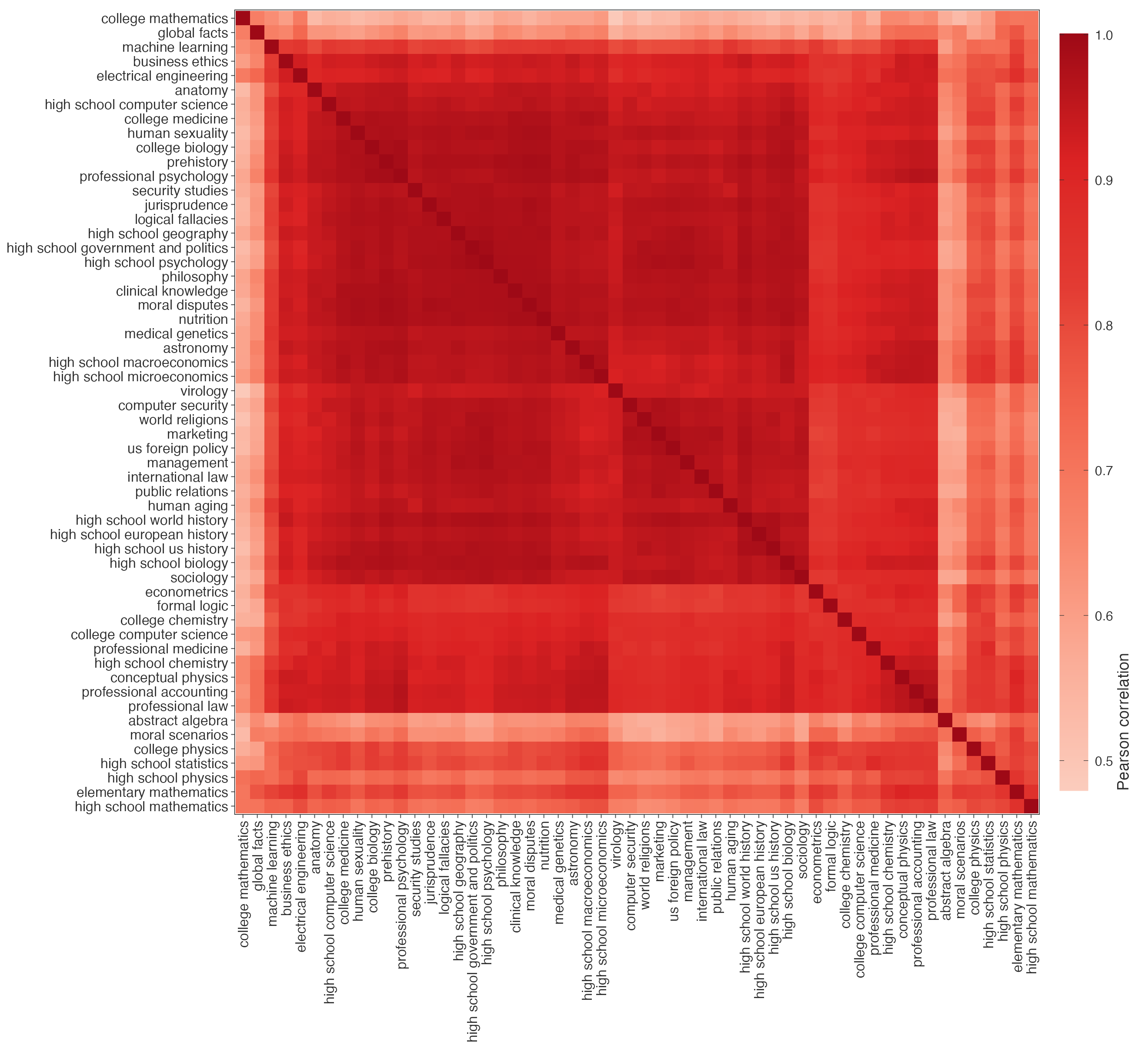}
    \caption{Heatmap of subject-level accuracy correlations across models. For each pair of MMLU subjects, the Pearson correlation between model-level subject accuracies is computed. Subjects are ordered by hierarchical clustering.}
    \label{fig:mmlu-subject-correlation}
\end{figure}

The data reveal several notable patterns. First, item counts exhibit high heterogeneity across domains. As illustrated in Figure~\ref{fig:mmlu-subject-item-counts}, the number of items per subject across the 56 MMLU domains ranges from 91 to 1,534, with a median of 173. The largest two domains are professional law and moral scenarios, with 1,534 and 895 items, respectively. Conversely, the smallest two domains are college physics and US foreign policy, with 91 and 99 items, respectively. 
Second, psychometric properties vary substantially across domains. For example, Figure~\ref{fig:mmlu-item-accuracy-by-subject} presents boxplots of item-level mean accuracy, defined as the proportion of models that answer each item correctly. Overall accuracy is high in domains such as marketing and high school government and politics, where item-level mean accuracy exceeds $50\%$ for all items. In contrast, it drops considerably in math-related subjects such as high school mathematics and abstract algebra, with mean accuracy falling below $50\%$ on most items. Third, LLM performance is positively correlated across different domains; models that perform well in one domain generally excel in others, and vice versa. Figure~\ref{fig:mmlu-subject-correlation} gives the heat map of a $56 \times 56$ correlation matrix for the 56 domains, where each off-diagonal entry of the matrix records the correlation between the raw score vectors for two domains.  
The between-domain correlations range between 0.489 and 
0.991, with a median of 0.916.

These data patterns pose significant challenges to standard measurement methods when addressing the questions above. First, consider evaluating an LLM's domain-specific capabilities. Two scoring approaches are commonly used: calculating a domain-specific total score or fitting a unidimensional item response theory (IRT) model, such as the two-parameter logistic (2PL) model \citep{birnbaum1968some}, separately for each domain. However, because both methods generate scores independently for different domains, they fail to borrow information from the other domains when producing scores for one domain. Consequently, the resulting estimates may not be accurate enough, particularly for domains with a limited number of items. Yet, as shown in Figure~\ref{fig:mmlu-subject-correlation}, raw scores across different domains are highly correlated. This suggests that a significant amount of information could be leveraged from related domains to improve the measurement accuracy of any single domain.

Second, consider deriving an aggregate score to summarize the overall performance of LLMs. While leaderboards commonly calculate total or weighted correct scores across all domains, this approach faces two limitations. First, substantial differences in domain lengths (see Figure~\ref{fig:mmlu-subject-item-counts}) mean raw totals fail to provide an equitable summary. Second, high variability in psychometric properties across domains (see Figure~\ref{fig:mmlu-item-accuracy-by-subject}) makes it difficult to assign appropriate weights. Alternatively, one could apply a unidimensional IRT model to the pooled response matrix (i.e., the matrix of LLMs' responses to items across all domains). However, this approach risks producing a biased score dominated by domains with larger item counts.

\section{Proposed Method}\label{sec:method}

\subsection{An IRT Model for Multi-domain LLM Evaluation}

We consider $i=1, \ldots, N$ LLMs being evaluated on $t = 1, \ldots, T$ benchmark domains. Each domain $t$ contains $J_t$ items. For each LLM $i$ and item $j$ in domain $t$, we observe $y_{itj} \in \{0, 1\}$, where $y_{itj} = 1$ if LLM $i$ answers the item correctly, and 0 otherwise. In what follows,  we introduce a multi-domain IRT model for the LLM performance data.

\paragraph{Multi-domain IRT Model.}
Assume that each LLM is associated with a $T$-dimensional parameter vector $ (\theta_{i1}, \theta_{i2}, \ldots, \theta_{iT})^\top$, with $\theta_{it}$ characterizing the capability of LLM $i$ on the $t$th domain. We further assume that the performance of LLM $i$ on the $j$th item in the $t$th domain is determined only by $\theta_{it}$, as well as some item-specific parameters to be defined in the sequel. More specifically, we assume the item response probability for $y_{itj}$ to follow a two-parameter logistic (2PL) model \citep{birnbaum1968some}, one of the most commonly applied IRT models \citep{chen2025item}. That is, 
\begin{equation}\label{eq:2pl}
P(y_{itj}=1\vert \theta_{it}, a_{tj}, d_{tj}) = \frac{\exp(a_{tj} \theta_{it} + d_{tj})}{1+\exp(a_{tj} \theta_{it} + d_{tj})},
\end{equation}
where $a_{tj} \geq 0$ is known as the discrimination parameter and $d_{tj}$ is known as the easiness parameter. Under this model assumption, for a given item $j$ in domain $t$, the larger the value of capability trait $\theta_{it}$, the higher the chance that the LLM is able to answer it correctly. We impose the constraints that for each $t = 1, \ldots, T$, 
\begin{equation}\label{eq:constr1}
    \sum_{i=1}^N \theta_{it} = 0 \mbox{~~~and ~}    \left(\sum_{i=1}^N \theta_{it}^2\right)/N = 1
\end{equation} 
to fix the indeterminacy in the location and scaling of the capability traits; otherwise, one can apply transformations to the capability traits and the item-specific parameters without changing the distribution of $y_{itj}$. 
Here, $\theta_{it}$, $a_{tj}$, and $d_{tj}$ are treated as unknown parameters.

\paragraph{General Intelligence Trait and Cross-domain Heterogeneity.}
We shall note that the multi-domain IRT model can be seen as a multidimensional IRT model with $T$ latent traits. However, it differs from the traditional model specification \citep[see, e.g.,][]{reckase2009multidimensional}, where $(\theta_{i1}, \theta_{i2}, \ldots, \theta_{iT})^\top$, $i=1, \ldots, N$, are independent and identically distributed random vectors, typically assumed to follow a multivariate normal distribution. We choose the current model specification for its flexibility in accommodating complex capability trait distributions (e.g., skewness, clustering, and multimodality).
On the other hand, we shall note that the lack of a distributional assumption makes it more challenging to characterize the dependence, or equivalently, the heterogeneity, of the capability traits and to further define the general intelligence trait as a summary of the $T$ traits. In what follows, we describe the proposed approach to achieving these objectives.
 
We first define the general intelligence trait. Let $\theta_{it}^*$ be the true latent trait level, satisfying the same constraints in \eqref{eq:constr1}, and let $a_{tj}^* \geq 0$ and $d_{tj}^*$ be the true item-specific parameters. Denote $\boldsymbol{\theta}^*_{t} = (\theta_{1t}^*, \ldots, \theta_{Nt}^*)^\top \in \mathbb R^N$ as the trait vector for the $t$th domain. For any given vector $\boldsymbol{\theta} \in \mathbb R^N$ satisfying the same constraints in \eqref{eq:constr1}, the sine of the angle between $\boldsymbol{\theta}$ and $\boldsymbol{\theta}^*_{t}$ is defined as 
$\sin(\boldsymbol{\theta}, \boldsymbol{\theta}^*_{t} ) = \sqrt{1 - \left({\boldsymbol{\theta}^\top \boldsymbol{\theta}^*_{t} }/{N}\right)^2}.$
When the two vectors follow similar directions in $\mathbb R^N$, the sine function becomes smaller and takes value zero when the two vectors are identical up to sign. We define the general intelligence trait as the vector in $\mathbb R^N$ that is on average closest in direction to all $\boldsymbol{\theta}^*_{1}$, \ldots, $\boldsymbol{\theta}^*_{T}$. More specifically,  we define 
\begin{equation}\label{eq:general vector}
\boldsymbol{\theta}^*_{G} = \argmin_{\boldsymbol{\theta} \in \mathbb R^N} \sum_{t=1}^T \sin(\boldsymbol{\theta}, \boldsymbol{\theta}^*_{t} ), \mbox{~s.t.~} \sum_{i=1}^N \theta_{i} = 0 ,   \left(\sum_{i=1}^N \theta_{i}^2\right)/N = 1,\mbox{~and~}\boldsymbol{\theta}^\top
\left(
\sum_{t=1}^T\boldsymbol{\theta}_t^*
\right)\geq 0,    
\end{equation}
where we assume the minimizer exists and is unique. The last constraint is essential to ensure the uniqueness and interpretation of 
the general intelligence trait $\boldsymbol{\theta}^*_{G}$. Without this constraint, $- \boldsymbol{\theta}^*_{G}$ would also be a solution to the constrained optimization problem.

The value $\min_{\boldsymbol{\theta}} T^{-1}\sum_{t=1}^T \sin(\boldsymbol{\theta}, \boldsymbol{\theta}^*_{t})$ naturally provides a measure of cross-domain heterogeneity. In settings where the cross-domain heterogeneity is low, as in many AI benchmarking applications,  
this value is expected to be small. In what follows, we introduce an estimator for $\boldsymbol{\theta}_t$, $t=1, \ldots, T$, and $\boldsymbol{\theta}_G$, in which the prior information about the low cross-domain heterogeneity is incorporated through a penalty term based on this measure of heterogeneity.

\subsection{Proposed Estimator}\label{subsec:proposed-estimators}

We propose a regularized estimator that jointly estimates the
domain-specific capability traits and the general intelligence trait, with the prior knowledge that the cross-domain heterogeneity is low. Recall the notation $\boldsymbol{\theta}_t
= (\theta_{1t},\ldots,\theta_{Nt})^\top$. We further use
$\boldsymbol{a}_t
=
(a_{t1},\ldots,a_{tJ_t})^\top$
and 
$\boldsymbol{d}_t
=
(d_{t1},\ldots,d_{tJ_t})^\top$
to simplify the notation for the item-specific parameters for the $t$th domain. 
The negative log-likelihood contribution from the $t$th domain is 
$$L_t\left(
\boldsymbol{\theta}_t,
\boldsymbol{a}_t,
\boldsymbol{d}_t
\right)
=
- \sum_{i=1}^N\sum_{j=1}^{J_t}
\bigg[
y_{itj}\left(a_{tj}\theta_{it}+d_{tj}\right)
- \log(1+\exp(a_{tj}\theta_{it}+d_{tj})) \bigg].$$

We propose the following regularized estimator, with tuning parameter \(\lambda\in (0,\infty)\) and constraint constant $C > 0$:
\begin{equation}
\begin{aligned}
\left(
\left\{
\widehat{\boldsymbol{\theta}}_t^\lambda,
\widehat{\boldsymbol{a}}_t^\lambda,
\widehat{\boldsymbol{d}}_t^\lambda
\right\}_{t=1}^T,
\widehat{\boldsymbol{\theta}}_G^\lambda
\right) ~\in~ 
\argmin_{\left\{
{\boldsymbol{\theta}}_t,
{\boldsymbol{a}}_t,
{\boldsymbol{d}}_t
\right\}_{t=1}^T,
{\boldsymbol{\theta}}}
&~~ \left(\sum_{t=1}^T
\frac{1}{J_t}L_t\left(
\boldsymbol{\theta}_t,
\boldsymbol{a}_t,
\boldsymbol{d}_t
\right)\right)
+
\lambda
\left (\sum_{t=1}^T \sin\left(
\boldsymbol{\theta},
\boldsymbol{\theta}_t
\right)
\right),\\
s.t.    & ~~ \sum_{i=1}^N \theta_{it} = 0,   \left(\sum_{i=1}^N \theta_{it}^2\right)/N = 1, |\theta_{it}| \leq C, \\
& ~~ 0 \leq a_{tj} \leq C, |d_{tj}| \leq C,\\
&~~ \sum_{i=1}^N \theta_{i} = 0,   \left(\sum_{i=1}^N \theta_{i}^2\right)/N = 1, \boldsymbol{\theta}^\top\sum_{t=1}^T\boldsymbol{\theta}_t
\geq 0, \\
&~~i = 1, \ldots, N, t = 1, \ldots, T, j = 1, \ldots, J_t. 
\end{aligned}
\label{eq:proposed-regularized-estimator}
\end{equation}
The objective function involves two terms. The first term $\left(\sum_{t=1}^T
\frac{1}{J_t}L_t\left(
\boldsymbol{\theta}_t,
\boldsymbol{a}_t,
\boldsymbol{d}_t
\right)\right)$ is the loss function, which is a negative joint log-likelihood function for the multi-domain IRT model weighted by the numbers of items of different domains, and the second term $\lambda
\left (\sum_{t=1}^T \sin\left(
\boldsymbol{\theta},
\boldsymbol{\theta}_t
\right)\right)$ is a regularization term that penalizes solutions with high cross-domain heterogeneity. {The weight $1/J_t$ in the loss function normalizes the domain-specific loss by the number of items, so that domains with more items do not disproportionately influence the estimation of the general intelligence trait. This weighting places different domains on a comparable scale and ensures that the relative contribution of each domain, as well as the interpretation of the regularization parameter $\lambda$, is less affected by the domain sizes.}   The tuning parameter $\lambda$ balances the trade-off between the model fit determined by the first term and the prior knowledge on cross-domain heterogeneity in the second term. This parameter controls the degree of information sharing across domains. A smaller
value of $\lambda$ places more emphasis on the domain-specific likelihoods, whereas a larger value encourages
stronger pooling through the general intelligence trait. Thus, increasing $\lambda$ can reduce estimation
variance when the domains are similar, but may introduce additional bias when the domain-specific
traits are heterogeneous. As shown in Section~\ref{sec:theory}, the optimal choice of $\lambda$ depends on $N$ and $\{J_t\}_{t=1}^T$, under which both the estimation of the general intelligence trait and the domain-specific traits have the optimal rate. In practice, we choose $\lambda$ via a cross-validation procedure, as detailed in Section B.1 of the Supplementary Materials.   

We further comment on the constraints in \eqref{eq:proposed-regularized-estimator}. First, the standard normalization constraints in \eqref{eq:constr1} are imposed on the domain-specific traits, and similar constraints are also imposed on the general intelligence trait. Second, the constraint $\boldsymbol{\theta}^\top
\sum_{t=1}^T\boldsymbol{\theta}_t
\geq 0$ is the empirical version of the one involved in the definition of $\boldsymbol{\theta}_G^*$. Third, we constrain $a_{tj}$ to be nonnegative to ensure the monotone relationship between the domain-specific traits and item response performance. Such nonnegative constraints are common when applying the IRT model to educational testing \citep[e.g.,][]{chen2025item}. Finally, we note that the constraints involving the constant $C$ are standard in high-dimensional GLFM analysis \citep[see, e.g., ][]{chen2020structured,chen2022determining}, which ensures that an incoherence condition holds for low-rank matrix estimation. Consistent with observations in the GLFM literature, we also find that the estimation results are not sensitive to the choice of $C$. In our implementation in Sections \ref{sec:simulation} and \ref{sec:real}, we set $C = 4$.

{For computation, we minimize the objective
function in \eqref{eq:proposed-regularized-estimator} using a gradient-based adaptive moment estimation algorithm combined with alternating projections
\citep{kingma2015adam,lewis2009local}. At each
iteration, we first apply a gradient-based update to all parameters and then
project them back onto their respective constraint sets. The item
parameters are projected directly onto their box constraints. For each
domain-specific trait $\boldsymbol{\theta}_t$, we alternate between
centering and normalization and projection onto the entrywise bounded
set until all constraints are satisfied up to a prescribed tolerance.
The general intelligence trait is similarly centered and normalized,
with an additional sign adjustment to satisfy the constraint $\boldsymbol{\theta}^\top\sum_{t=1}^T\boldsymbol{\theta}_t
\geq 0$.  See Section B.3 of the Supplementary Materials for further details.}

We conclude this section by discussing two special cases of the proposed estimator. In the first case, we let $\lambda$ converge to zero, so that no penalty on the cross-domain heterogeneity is imposed. In this case, $\left(
\widehat{\boldsymbol{\theta}}_t^0,
\widehat{\boldsymbol{a}}_t^0,
\widehat{\boldsymbol{d}}_t^0
\right)$ is completely determined by $L_t\left(
\boldsymbol{\theta}_t,
\boldsymbol{a}_t,
\boldsymbol{d}_t
\right)$, which does not borrow information from the other domains and thus may suffer from high variance. The second case is when $\lambda$ diverges to infinity. In this case, the penalty term leads to  
$\hat{\boldsymbol{\theta}}_1^{\infty} = \cdots = \hat{\boldsymbol{\theta}}_T^{\infty} = \hat{\boldsymbol{\theta}}_G^{\infty},$
with 
$\hat{\boldsymbol{\theta}}_G^{\infty}$ minimizing the negative weighted joint log-likelihood
$\sum_{t=1}^T
(L_t\left(
\boldsymbol{\theta}_t,
\boldsymbol{a}_t,
\boldsymbol{d}_t
\right)/J_t)$. This estimator ignores cross-domain differences and thus may suffer from the bias they introduce. We will formalize these intuitions in Section~\ref{sec:theory} below.

\section{Theoretical Properties}\label{sec:theory}

We establish the theoretical properties for the proposed estimator. We first introduce notation and define quantities needed for the theory. For two real numbers $a$ and $b$, we denote their minimum by $\min \{a, b\}$ or $a \wedge b$ and their maximum by $\max \{a, b\}$ or $a \vee b$, respectively. For two positive real sequences $\left\{a_n\right\}_{n=1}^{\infty}$ and $\left\{b_n\right\}_{n=1}^{\infty}, a_n \lesssim b_n$ or $b_n \gtrsim a_n$ means there exists a universal constant $C>0$ such that $a_n \leq C b_n$ for all $n$. We say 
$a_n \asymp b_n$ if $a_n \lesssim b_n$ and $a_n \gtrsim b_n$. 
For a vector $\boldsymbol{u}$, $\|\boldsymbol{u}\|_2$ and $\|\boldsymbol{u}\|_\infty$ stand for its Euclidean and max norms. For a matrix $\boldsymbol{A},\|\boldsymbol{A}\|_2$, $\|\boldsymbol{A}\|_{\mathrm{F}}$, and $\|\boldsymbol{A}\|_{\mathrm{\text{max}}}$ represent its spectral, Frobenius, and max norms, respectively. 
For each domain $t$, we denote 
$m_{itj}^* = a_{tj}^*\theta_{it}^*+d_{tj}^*$ as the true log-odds for the item response $y_{itj}$ and let 
$\boldsymbol{M}_t^* = (m_{itj}^*)_{i = 1, \ldots, N, j = 1, \ldots, J_t}$ be the $N$ by $J_t$ log-odds matrix for the $t$th domain.   
Let $J=\sum_{t=1}^T J_t$ denote the total number of items across all domains. For $a<b\in \mathbb{R}$, $\text{Uniform}(a,b)$ denotes uniform distribution on $[a,b]$. We further define the aggregated log-odds matrix  
$\boldsymbol{M}^*
=
\left[
\boldsymbol{M}_1^*,
\ldots,
\boldsymbol{M}_T^*
\right] 
\in\mathbb{R}^{N\times J},$
which combines $\boldsymbol M_1^*$, $\ldots$, $\boldsymbol M_T^*$ by column.

We further introduce an alternative measure of domain heterogeneity, defined by 
$h
=
\max_{1\leq t\leq T}
\sin\!\left(
\boldsymbol{\theta}_{G}^{*},
\boldsymbol{\theta}_{t}^{*}
\right),$
i.e., the largest angle between $\boldsymbol{\theta}^*_{G}$ and $\boldsymbol{\theta}_{t}^{*}$, which naturally characterizes the estimation bias when the tuning parameter $\lambda$ is over-selected. 
A small $h$ means the capability of LLMs in different domains is more similar, as in this case, $\boldsymbol{\theta}_{t}^{*}$s are clustered together near $\boldsymbol{\theta}^*_{G}$. 
The case $h = 0$ reduces to the setting where all domain-specific capability vectors coincide with the general capability vector. Throughout the paper, we assume that $h<1$, i.e., there does not exist a domain-specific trait that is orthogonal to the general intelligence trait. 
Theorem~\ref{thm:proposed-common-direction} below gives an upper bound for the estimation of domain-specific capability traits under the optimal choice of the tuning parameter $\lambda$, established under Assumptions \ref{ass:signal strength} and \ref{ass:ThSV} below. 

\begin{assumption}
\label{ass:signal strength}
There exists a constant
$C>0$, such that $\max_{1\leq t\leq T}
\{\|\boldsymbol{\theta}_t^*\|_{\infty},\|\boldsymbol a_t^*\|_{\infty},\|\boldsymbol d_t^*\|_{\infty}\}
\leq C$.
Moreover, the discrimination parameters satisfy
$a_{tj}^*\geq 0$ for $t=1,\ldots, T$ and $j=1,\ldots,J_t$.
For every $t\in[T]$, the domain-specific capability trait satisfies $(\boldsymbol{\theta}_t^*)^\top\boldsymbol 1_N=0,
\|\boldsymbol{\theta}_t^*\|_2^2=N$.
There exists a constant $c>0$, independent of
$N$, $T$, and $\{J_t\}_{t=1}^T$, such that
$\min_{t\in[T]}\left\{
{
\|\boldsymbol a_t^*\|_2^2
}/{J_t}\right\}
\geq c.$
Moreover, we assume that for all $1\leq t\leq T$, $(\boldsymbol\theta_G^*)^\top\boldsymbol\theta_t^*\geq 0$, and $\|\boldsymbol\theta_G^*\|_{\infty}\leq C$.
\end{assumption}

\begin{assumption}\label{ass:ThSV}
  There exists a constant $C_p>0$, such that
$\max_{t\in[T]}J_t/\min_{t\in[T]}J_t
\leq C_p.
$
\end{assumption}

\begin{theorem}
\label{thm:proposed-common-direction}
Suppose Assumptions \ref{ass:signal strength} and \ref{ass:ThSV} hold. If we set $\lambda=\max_{1\leq t\leq T}c_{\lambda}\sqrt{N^2/(N\wedge J_t)}$ for a sufficiently large constant $c_{\lambda}$, there exists a constant $c_1>0$, such that, with probability at least $1-
c_1\sum_{t=1}^T
\left\{
\frac{1}{N+J_t}
+
\frac{1}{NJ_t}
\right\}$, for $t=1,\ldots,T$,
$$
\frac{\|\widehat{\boldsymbol{\theta}}_t^\lambda-\boldsymbol{\theta}_t^*\|_2}{\sqrt{N}} \lesssim \left[\sqrt{\frac{1}{N\wedge J}}+h\right]\wedge\sqrt{\frac{1}{N \wedge J_t}}. $$
\end{theorem}

Theorem~\ref{thm:proposed-common-direction} establishes upper bounds for the domain-specific capability traits. It means that on average, the entries of $\boldsymbol{\theta}_t^*$ are estimated at the rate of $\left[\sqrt{{1}/({N\wedge J})}+h\right]\wedge\sqrt{{1}/{(N \wedge J_t)}}$, where the two terms correspond to the convergence rates when $\lambda = \infty$ and $0$, respectively, as formalized in Proposition~\ref{prop:localglobal} below. 
 That is, $\sqrt{1/(N \wedge J_t)}$ is the estimation error for the trait of the $t$th domain when only using data from this domain (by setting $\lambda = 0$), without borrowing information from the other domains. $\sqrt{{1}/({N\wedge J}})+h$ is the estimation error when we fully borrow information across the domains by setting $\boldsymbol{\theta}_t$s to be identical (by setting $\lambda = \infty$). Here, $h$ is the bias term arising from ignoring across-domain heterogeneity, and $\sqrt{{1}/({N\wedge J})}$ {comes from the stochastic error in estimating the common capability direction using the pooled data across all domains.} 
 Theorem~\ref{thm:proposed-common-direction} suggests that the proposed estimator automatically adapts to the faster convergence rate between the two rates corresponding to $\lambda=0$ and $\lambda = \infty$. Moreover, a lower bound is further established for the estimation of $\boldsymbol{\theta}_t^*$, as given in Theorem 3 of the Supplementary Materials. Under certain asymptotic regimes, such as when $N \asymp J$ and $\{J_t\}_{t=1}^T$ are of the same order, the lower bound matches the upper bound in Theorem~\ref{thm:proposed-common-direction}, showing that the proposed estimator of the domain-specific traits is minimax optimal.

\begin{proposition}\label{prop:localglobal}
Suppose that Assumptions \ref{ass:signal strength} and
\ref{ass:ThSV} hold. Then there exists a constant $c_1>0$ such
that, with probability at least
$1-
c_1 \sum_{t=1}^T
\left\{
\frac{1}{N+J_t}
+
\frac{1}{NJ_t}
\right\}$, the following bound holds for all $t=1,\ldots,T$:
$${\|\widehat{\boldsymbol{\theta}}_t^0-\boldsymbol{\theta}_t^*\|_2}/{\sqrt{N}} \lesssim \sqrt{\frac{1}{N \wedge J_t}} 
\mbox{~~and~~}
{\|\widehat{\boldsymbol{\theta}}_t^\infty-\boldsymbol{\theta}_t^*\|_2}/{\sqrt{N}} \lesssim \sqrt{\frac{1}{N\wedge J}}+h. $$
\end{proposition}

We comment on Assumptions \ref{ass:signal strength} and 
\ref{ass:ThSV}. 
Assumption~\ref{ass:signal strength} is a signal strength requirement, which introduces regularity conditions on $\boldsymbol{\theta}_t^*$, $\boldsymbol a_t^*$, and $\boldsymbol d_t^*$. In fact, the uniform boundedness of $\boldsymbol{\theta}_t^*$, $\boldsymbol a_t^*$, and $\boldsymbol d_t^*$ implies $\|\boldsymbol M_t^*\|_{\text{max}}\leq C^2+C$, while $(\boldsymbol{\theta}_t^*)^\top\boldsymbol 1_N=0$ and $\|\boldsymbol{\theta}_t^*\|_2^2=N$ yield $\|\boldsymbol M_t^*\|_F^2=N(\|\boldsymbol a_t^*\|_2^2+\|\boldsymbol d_t^*\|_2^2)$. Therefore, the signal-strength condition implies, uniformly over $t\in[T]$,
$\sqrt{c}
\leq
{\|\boldsymbol M_t^*\|_F}/{\sqrt{NJ_t}}
\leq
\sqrt{2}\,C.$ The requirement of $(\boldsymbol\theta_G^*)^\top\boldsymbol\theta_t^*\geq 0$ ensures that the 
domain-specific capability traits are 
all positively associated with the 
general intelligence trait. As shown in Section B.2 of the Supplementary Materials, these assumptions hold with high probability when entries of $\boldsymbol{\theta}_t^*$, $\boldsymbol{a}_t^*$ and $\boldsymbol{d}_t^*$ are generated from sensible distributions. Assumption~\ref{ass:ThSV} imposes a mild balance condition on the
domain sizes, facilitating a unified theoretical analysis across
domains with heterogeneous numbers of items.

In Theorem \ref{thm:G factor rate}, we establish the convergence rate for the general intelligence trait $\boldsymbol{\theta}_G$.  

\begin{assumption}\label{ass:regularization}
The optimization

\begin{equation}\label{eq:pert}
\min_{\boldsymbol{\theta} \in \mathbb R^N} \sum_{t=1}^T \sin(\boldsymbol{\theta}, \boldsymbol{\theta}_{t} ), \mbox{~s.t.~} \sum_{i=1}^N \theta_{i} = 0 ,   \left(\sum_{i=1}^N \theta_{i}^2\right)/N = 1,\mbox{~and~}\boldsymbol{\theta}^\top
\left(
\sum_{t=1}^T\boldsymbol{\theta}_t
\right)\geq 0    
\end{equation}
has a unique solution for all $\boldsymbol\theta_t$ in a small neighborhood of $\boldsymbol\theta_t^*$, i.e., 
$\left\|\boldsymbol{\theta}_t-\boldsymbol{\theta}_t^*\right\|_2/\sqrt{N}<\epsilon$, for all $t=1, \ldots, T$ and a sufficiently small $\epsilon$. In addition, the minimizer of \eqref{eq:pert}, denoted by $\tilde{\boldsymbol{\theta}}_G$, as a function of $\boldsymbol\theta_t$, $t=1, \ldots, T$, satisfies 
${\Vert \tilde{\boldsymbol{\theta}}_G - {\boldsymbol{\theta}}_G^*\Vert_2} \lesssim \frac{1}{T} \sum_{t=1}^T  \Vert\boldsymbol{\theta}_t - \boldsymbol{\theta}_t^*\Vert_2.$
\end{assumption}

\begin{theorem}\label{thm:G factor rate}
Suppose that Assumptions \ref{ass:signal strength} and
\ref{ass:ThSV} hold. If we set $\lambda=\max_{1\leq t\leq T}\{c_{\lambda}\sqrt{N^2/(N\wedge J_t)}\}$ for a sufficiently large constant $c$, then with probability at least
$1-
c_1 \sum_{t=1}^T
\left\{
\frac{1}{N+J_t}
+
\frac{1}{NJ_t}
\right\}$,
$$\frac{\left\|\widehat{\boldsymbol{\theta}}_G^\lambda-\boldsymbol{\theta}_G^*\right\|_2}{\sqrt{N}} \lesssim \sqrt{\frac{1}{N\wedge J}}+h.$$
If, in addition, Assumption \ref{ass:regularization} holds, then 
$$\frac{\left\|\widehat{\boldsymbol{\theta}}_G^\lambda-\boldsymbol{\theta}_G^*\right\|_2}{\sqrt{N}} \lesssim \frac{1}{T}\sum_{t=1}^T\left\{\left[\sqrt{\frac{1}{N\wedge J}}+h\right] \wedge \sqrt\frac{1}{N\wedge J_t}\right\}.$$
\end{theorem}

Theorem \ref{thm:G factor rate} establishes upper bounds for the general intelligence trait. Specifically, the normalized estimation error
${\|\widehat{\boldsymbol{\theta}}_G^\lambda-\boldsymbol{\theta}_G^*\|_2}/{\sqrt{N}}$ is upper bounded by $\sqrt{1 /(N \wedge J)}+h$ under Assumptions \ref{ass:signal strength} and \ref{ass:ThSV}, which corresponds to the convergence rate when $\lambda=\infty$. This upper bound is further improved when an additional smoothness condition, Assumption~\ref{ass:regularization}, holds.
This assumption ensures that small perturbations in the domain-specific intelligence traits do not cause disproportionately large changes in the resulting general intelligence trait. {It holds when no domain-specific trait is exactly aligned with the general intelligence trait,
so that the objective function in \eqref{eq:pert} remains locally smooth, and when the domain-specific
traits are sufficiently dispersed around the general intelligence trait with moderate
cross-domain heterogeneity, so that the objective function retains sufficient local curvature.} Formal sufficient conditions that guarantee Assumption~\ref{ass:regularization} and additional justifications
are provided in
Propositions A.1 and A.2 of the Supplementary Materials.
When Assumption~\ref{ass:regularization} holds, the estimation error of 
$\widehat{\boldsymbol{\theta}}_G^\lambda$ also adapts to that of $\widehat{\boldsymbol{\theta}}_G^0$, defined as  the minimizer of the optimization
\begin{equation*}
\min_{\boldsymbol{\theta} \in \mathbb R^N} \sum_{t=1}^T \sin(\boldsymbol{\theta}, \widehat{\boldsymbol{\theta}}_t^{0} ), \mbox{~s.t.~} \sum_{i=1}^N \theta_{i} = 0 ,   \left(\sum_{i=1}^N \theta_{i}^2\right)/N = 1,\mbox{~and~}\boldsymbol{\theta}^\top
\left(
\sum_{t=1}^T\widehat{\boldsymbol{\theta}}_t^{0}
\right)\geq 0,    
\end{equation*}
which is obtained by pooling $\widehat{\boldsymbol{\theta}}_t^{0}$, $t=1, \ldots, T$. $\widehat{\boldsymbol{\theta}}_G^0$ has the following error bound.
\begin{proposition}\label{prop:localglobal2}
Suppose that Assumptions \ref{ass:signal strength} and
\ref{ass:ThSV} hold. Then there exists a constant $c_1>0$ such
that, with probability at least
$1-
c_1 \sum_{t=1}^T
\left\{
\frac{1}{N+J_t}
+
\frac{1}{NJ_t}
\right\}$, the following bound holds: ${\left\|\widehat{\boldsymbol{\theta}}_G^\infty-\boldsymbol{\theta}_G^*\right\|_2}/{\sqrt{N}} \lesssim \sqrt{\frac{1}{N\wedge J}}+h. $
If, in addition, Assumption \ref{ass:regularization} holds, then 
$$
{\left\|\widehat{\boldsymbol{\theta}}_G^0-\boldsymbol{\theta}_G^*\right\|_2}/{\sqrt{N}} \lesssim \frac{1}{T}\sum_{t=1}^T\sqrt{\frac{1}{N\wedge J_t}}. $$

\end{proposition}

Theorem \ref{thm:G factor rate} shows that $\widehat{\boldsymbol{\theta}}_G^{\lambda}$ adapts to the faster convergence rate between the two rates corresponding to $\lambda=\infty$ and $\lambda=0$, as
    $$
    \frac{1}{T} \sum_{t=1}^T\left\{\left[\sqrt{\frac{1}{N \wedge J}}+h\right] \wedge \sqrt{\frac{1}{N \wedge J_t}}\right\}\leq\left[\sqrt{\frac{1}{N \wedge J}}+h\right]\wedge \left\{\frac{1}{T} \sum_{t=1}^T \sqrt{\frac{1}{N \wedge J_t}}\right\}.
    $$

\section{Simulation Study}\label{sec:simulation}
We conduct a simulation study to evaluate the proposed estimator and compare it with alternatives when $\lambda = 0$ and $\infty$. We evaluate the performance of the three methods in estimating the general intelligence trait and domain-specific traits.  For the proposed method, we select the tuning parameter $\lambda$ by cross-validation. We then refit the model on the full dataset using the selected $\lambda$. See Section B.1 of the Supplementary Materials for details.

\subsection{Data Generation and Evaluation Criteria}
We consider $N = 100$ LLMs across $T = 18$ domains of varying sizes. Specifically, the number of items per domain, $J_t$, takes values in $\{50, 60, 70, 80, 90, 100\}$, with exactly 3 domains allocated to each size. 
To generate domain-specific capability vectors $\{\boldsymbol{\theta}_t^*\}_{t=1}^T$, we first generate a center vector $\bar{\boldsymbol{\theta}}$ that satisfies $(\bar{\boldsymbol{\theta}})^\top\boldsymbol{1}_N=0$ and $\|\bar{\boldsymbol{\theta}}\|_2^2=N$. Then, we construct $T$ mutually orthonormal vectors $\{\boldsymbol{v}_t\}_{t=1}^T$ satisfying $\boldsymbol{v}_t^\top\boldsymbol{1}_N=\boldsymbol{v}_t^\top\bar{\boldsymbol{\theta}}=0$ and $\|\boldsymbol{v}_t\|_2=1$, and set $\alpha_t=r_th_0$ for $1\leq t\leq T$, where $r_1=1$ and $\{r_t\}_{t=2}^T$ are independently sampled from  $\text{Uniform}(-1,1)$. The capability vectors are generated as 
$\boldsymbol{\theta}_{t}^*
=
\sqrt{1-\alpha_t^2}\bar{\boldsymbol{\theta}}
+
\sqrt{N}\alpha_t\boldsymbol{v}_t,\ \text{for}\ 1\leq t\leq T.$
This construction ensures that $(\boldsymbol{\theta}_t^*)^\top\boldsymbol{1}_N=0$, $\|\boldsymbol{\theta}_t^*\|_2^2=N$, and $\max_{t\in[T]}\sin(\bar{\boldsymbol{\theta}},\boldsymbol{\theta}_t^*)=h_0$. The general intelligence trait $\boldsymbol{\theta}_G^*$ is obtained by solving the optimization (\ref{eq:general vector}). Typically, $\boldsymbol{\theta}_G^*$ is not exactly the same as the center vector $\bar{\boldsymbol{\theta}}$, and thus, $h$ is smaller than $h_0$. {Nevertheless, as $N,T \to \infty$, $\boldsymbol{\theta}_G^*$ converges to $\bar{\boldsymbol{\theta}}$ and $h$ converges to $h_0$, making $h_0$ a good proxy for $h$.} As it is difficult to precisely control $h$ in the data generation, we instead control $h_0$ to take values in $\{0.1,0.2,\ldots,0.5\}$ to measure 
 the level of heterogeneity between domains. 
For item parameters, independently across domains and items, we generate $a_{tj}^*$ from $\operatorname{Uniform}(0.5,1.5)$ and $d_{tj}^*$ from $\operatorname{Uniform}(-1,1)$, and sample $y_{itj}$ from the 2PL model \eqref{eq:2pl} for $1\leq t\leq T$, $1\leq i\leq N$ and $1\leq j\leq J_t$. For each value of $h_0$, we run 100 independent replications.

We report the following performance metrics. 
For the general intelligence trait, we compute the estimation error based on the Euclidean distance between $\widehat{\boldsymbol{\theta}}_G$ and $\boldsymbol{\theta}_G^*$,  
${\|\widehat{\boldsymbol{\theta}}_G-\boldsymbol{\theta}_G^*\|_2}/{\sqrt{N}}$. Similarly, for each domain-specific trait, we calculate
${\|\widehat{\boldsymbol{\theta}}_t-\boldsymbol{\theta}_t^*\|_2}/{\sqrt{N}}$, for $1\leq t\leq T$.

\subsection{Results}

\begin{figure}[htbp]
    \centering
    \includegraphics[width=\textwidth]{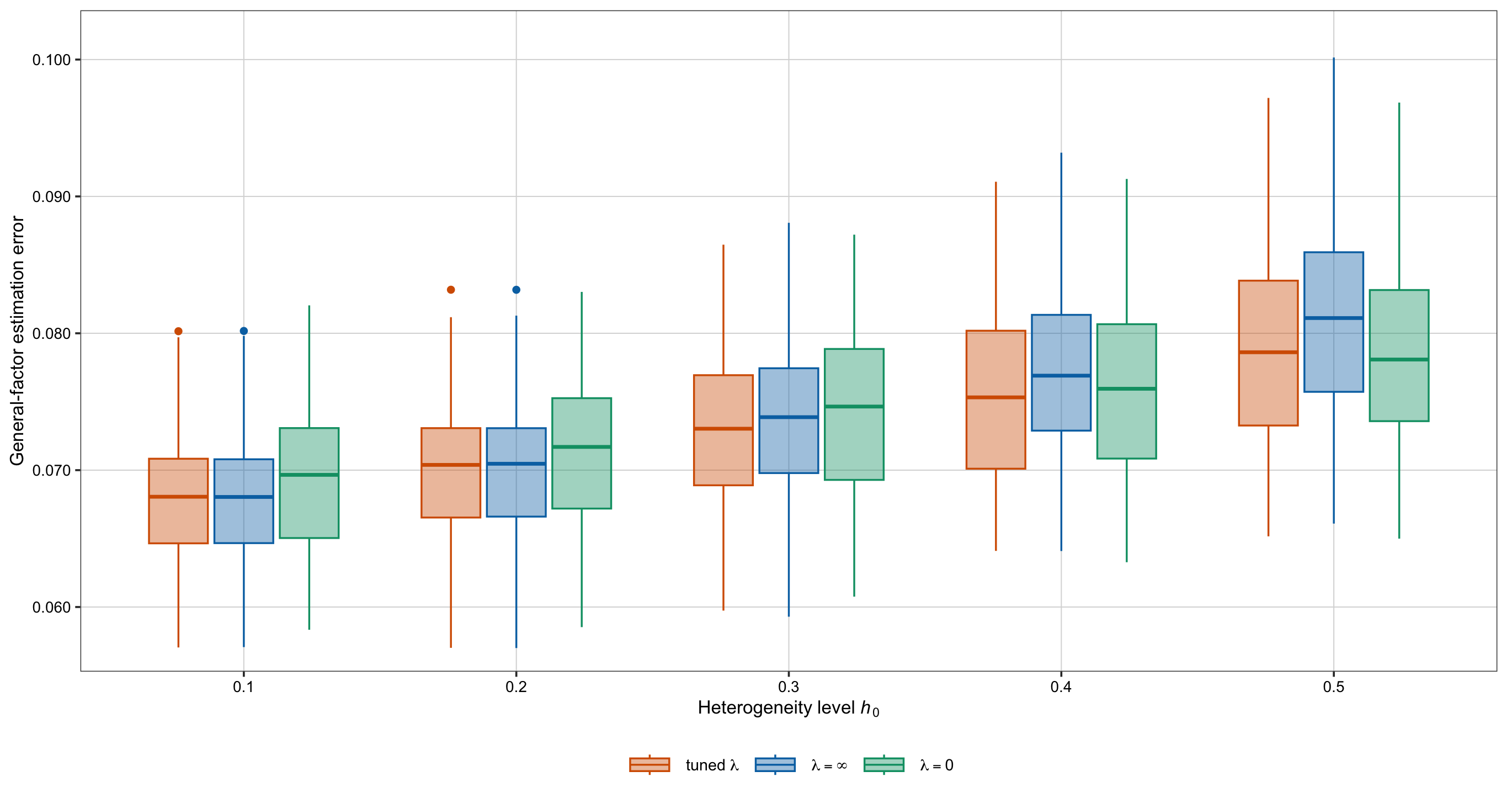}
    \caption{Box plots of the general intelligence trait estimation error ${\|\widehat{\boldsymbol{\theta}}_G-\boldsymbol{\theta}_G^*\|_2}/{\sqrt{N}}$ across replications under different levels of heterogeneity $h_0$.
    }
    \label{fig:simulation_general_error}
\end{figure}

\begin{figure}[htbp]
    \centering
    \includegraphics[width=\textwidth]{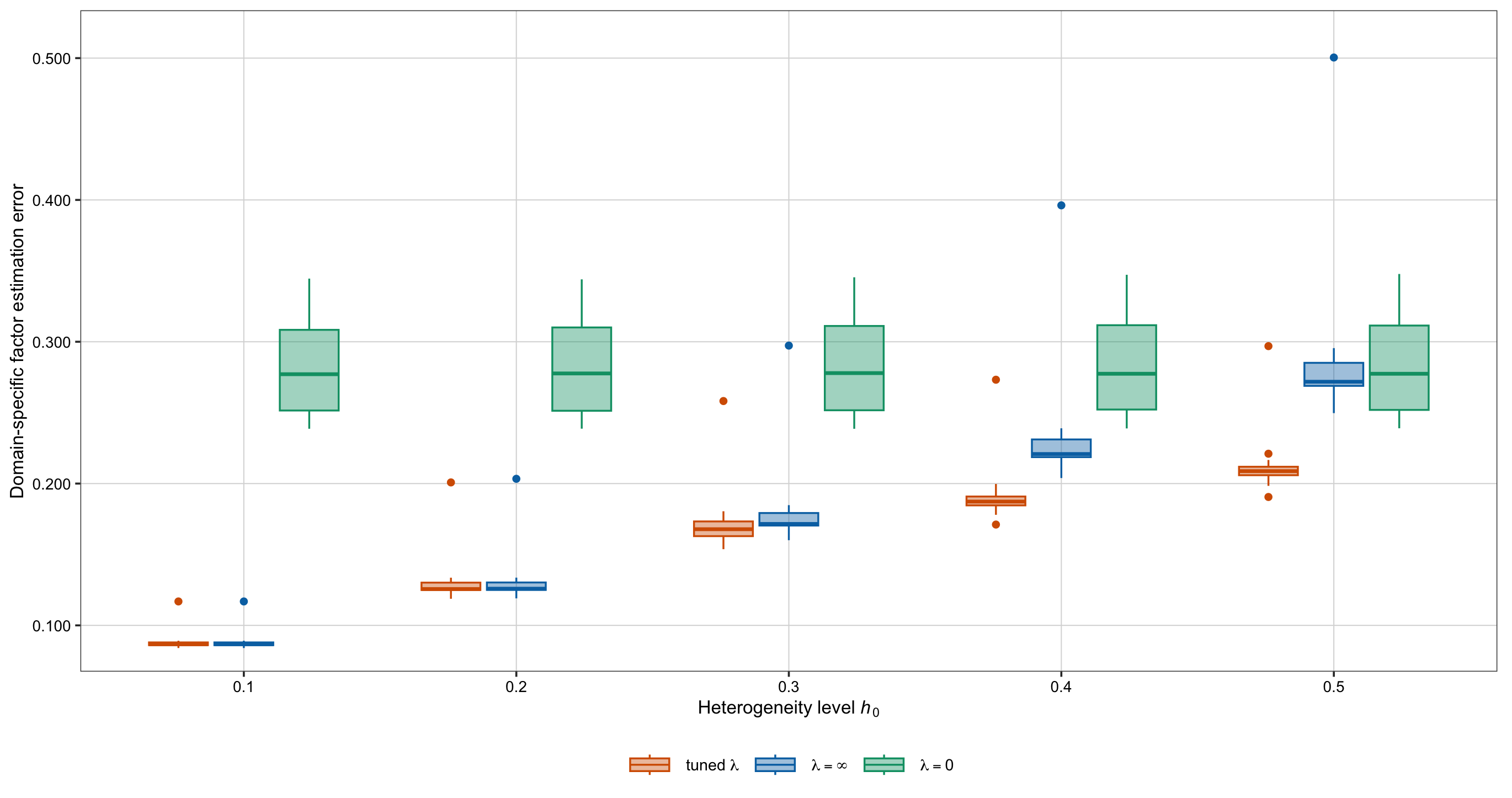}
    \caption{Box plots across the 18 domains of the replication-averaged domain-specific error $\{{\|\widehat{\boldsymbol{\theta}}_t-\boldsymbol{\theta}_t^*\|_2}/{\sqrt{N}}\}_{t=1}^T$ across domains under different levels of heterogeneity $h_0$.
    }
    \label{fig:simulation_domain_error}
\end{figure}

The results are given in Figures \ref{fig:simulation_general_error} and \ref{fig:simulation_domain_error}.  
Figure~\ref{fig:simulation_general_error} compares the three estimators in estimating the general intelligence trait for settings corresponding to different heterogeneity levels. Specifically, for each value of $h_0$ and each estimator, Figure~\ref{fig:simulation_general_error} gives box plots of ${\|\widehat{\boldsymbol{\theta}}_G-\boldsymbol{\theta}_G^*\|_2}/{\sqrt{N}}$ over 100 independent simulations. The proposed method achieves the best overall performance across all levels of heterogeneity. That is, when the heterogeneity level is relatively low, i.e., when $h_0 = 0.1$ and $0.2$, $\widehat{\boldsymbol{\theta}}_G^{0}$ tends to have the worst performance among the three estimators.
This is due to the higher variance of this estimator. Under these settings, the proposed 
estimator and $\widehat{\boldsymbol{\theta}}_G^{\infty}$ perform similarly. 
When the heterogeneity level is relatively high, i.e., when  $h_0 = 0.4$ and $0.5$, 
$\widehat{\boldsymbol{\theta}}_G^{\infty}$ tends to perform the worst, due to its higher bias. In these cases, the proposed estimator and $\widehat{\boldsymbol{\theta}}_G^{0}$
perform similarly. 
For a median level of heterogeneity, i.e., $h_0 = 0.3$, the proposed estimator tends to slightly outperform both $\widehat{\boldsymbol{\theta}}_G^{0}$ and
$\widehat{\boldsymbol{\theta}}_G^{\infty}$. 
However, we note that the bias and variance in estimating the general intelligence trait are low under these settings; thus, the differences among the three estimators are not substantial.

Figure~\ref{fig:simulation_domain_error} reports 
the errors regarding the estimation of the 
domain-specific traits. For each domain $t$, each estimator,  and each value of $h_0$, the error metric ${\|\widehat{\boldsymbol{\theta}}_t-\boldsymbol{\theta}_t^*\|_2}/{\sqrt{N}}$ is averaged over 100 independent replications. Then, for each estimator and each value of $h_0$, a box plot is created based on the averaged error metrics across the $T$ domains. Again, the proposed estimator shows the best overall performance across levels of heterogeneity. Specifically, the error metric of $\widehat{\boldsymbol{\theta}}_t^{0}$ tends to be stable across different values of $h_0$, as this estimator does not borrow information across domains and thus is not affected by the cross-domain heterogeneity. This estimator performs substantially worse than the other two when cross-domain heterogeneity is relatively low, making it beneficial to borrow information across domains. In fact, under the current simulation settings, $\widehat{\boldsymbol{\theta}}_t^{0}$ performs substantially worse than the proposed estimator for all  $h_0 = 0.1$, \ldots, $0.5$. We further compare 
$\widehat{\boldsymbol{\theta}}_t^{\infty}$ and the proposed estimator. When  $h_0 = 0.1$ and 0.2, the two estimators perform similarly. When $h_0 = 0.3, 0.4$ and 0.5, the proposed estimator substantially outperforms $\widehat{\boldsymbol{\theta}}_t^{\infty}$, as the latter starts to suffer from the high bias brought by over-borrowing information. 

\section{Application to MMLU Data}\label{sec:real}

We apply the proposed framework to the MMLU response data and address the two research questions introduced in Section~\ref{sec:data}. The proposed estimator jointly produces the domain-specific traits \(\{\widehat{\boldsymbol{\theta}}_{t}^{\lambda}\}_{t=1}^{T}\), the general intelligence trait \(\widehat{\boldsymbol{\theta}}_{G}^{\lambda}\), and the item parameters \(\{(\widehat{\boldsymbol a}_{t}^{\lambda},\widehat{\boldsymbol d}_{t}^{\lambda})\}_{t=1}^{T}\). Throughout the analysis, five-fold cross-validation selects a penalty parameter \(\widehat\lambda=156.74\); detailed results are reported in the Supplementary Materials C.2. We start with the first research question by examining the estimated domain-specific traits. We then turn to the second question by analyzing the estimated general intelligence trait and comparing the resulting overall ranking against two benchmark summaries. Finally, to further validate our results, we examine the relationship between LLM model size and the estimated general intelligence trait and domain-specific capability traits, which is presented in Section C.3 of the Supplementary Materials.

\subsection{Domain-Specific Capability Measurement}

We begin with the first research question, which concerns the measurement of domain-specific capability. The main motivation for the proposed estimator is that LLM performance is highly correlated across subjects, so subject-level capability need not be estimated in complete isolation. The subject-by-subject estimated traits and ranking results are provided in the code repository. Table~\ref{tab:subject-rank-vs-general-rank} summarizes how each estimated domain-specific trait \(\widehat{\boldsymbol{\theta}}_{t}\) relates to the estimated general intelligence trait \(\widehat{\boldsymbol{\theta}}_{G}\). We report two measures. The first is the sine discrepancy
\(
\sin\!\left(\widehat{\boldsymbol{\theta}}_{t}, \widehat{\boldsymbol{\theta}}_{G}\right),
\)
which is the geometric quantity used in constructing the estimator. The value of this measure is close to zero when the two vectors are nearly collinear. The second is the Spearman rank correlation between the subject-specific ranking and the ranking induced by \(\widehat{\boldsymbol{\theta}}_{G}\). Across the 56 retained subjects, the sine discrepancy ranges from nearly \(0\) to \(0.749\), with a median of \(0.106\). Thus, for most subjects, the fitted subject-specific trait remains close to the dominant general direction.

\begin{table}[!h]
\centering
\scriptsize
\caption{\(\sin\!\left(\widehat{\boldsymbol{\theta}}_{t},\widehat{\boldsymbol{\theta}}_{G}\right)\) and Spearman's \(\rho\) between each subject-specific ranking and the ranking induced by the estimated general intelligence trait \(\widehat{\boldsymbol{\theta}}_{G}\). Subjects are ordered by sine discrepancy from small to large.}
\label{tab:subject-rank-vs-general-rank}
\begin{tabular}{lcc|lcc}
\toprule
Subject & \(\sin\!\left(\widehat{\boldsymbol{\theta}}_{t},\widehat{\boldsymbol{\theta}}_{G}\right)\) & Spearman \(\rho\) & Subject & \(\sin\!\left(\widehat{\boldsymbol{\theta}}_{t},\widehat{\boldsymbol{\theta}}_{G}\right)\) & Spearman \(\rho\) \\
\midrule
professional psychology & 0.000 & 1.000 & anatomy & 0.108 & 0.986 \\
prehistory & 0.001 & 1.000 & high school us history & 0.108 & 0.992 \\
clinical knowledge & 0.001 & 1.000 & virology & 0.109 & 0.981 \\
nutrition & 0.012 & 1.000 & marketing & 0.111 & 0.994 \\
moral disputes & 0.016 & 1.000 & business ethics & 0.114 & 0.977 \\
philosophy & 0.031 & 0.999 & computer security & 0.115 & 0.981 \\
high school psychology & 0.034 & 1.000 & security studies & 0.120 & 0.991 \\
high school geography & 0.035 & 0.998 & international law & 0.120 & 0.988 \\
high school macroeconomics & 0.037 & 0.999 & world religions & 0.128 & 0.987 \\
high school government and politics & 0.045 & 0.998 & high school european history & 0.133 & 0.986 \\
college medicine & 0.049 & 0.997 & professional medicine & 0.135 & 0.988 \\
high school microeconomics & 0.053 & 0.997 & public relations & 0.142 & 0.981 \\
college biology & 0.056 & 0.997 & sociology & 0.149 & 0.983 \\
high school world history & 0.058 & 0.996 & electrical engineering & 0.168 & 0.962 \\
logical fallacies & 0.060 & 0.996 & college computer science & 0.170 & 0.972 \\
professional accounting & 0.062 & 0.995 & college chemistry & 0.180 & 0.964 \\
jurisprudence & 0.062 & 0.994 & econometrics & 0.191 & 0.968 \\
human sexuality & 0.075 & 0.991 & formal logic & 0.252 & 0.961 \\
professional law & 0.080 & 0.993 & high school statistics & 0.281 & 0.946 \\
medical genetics & 0.082 & 0.986 & machine learning & 0.319 & 0.847 \\
high school biology & 0.088 & 0.996 & college physics & 0.321 & 0.928 \\
high school computer science & 0.091 & 0.992 & elementary mathematics & 0.361 & 0.936 \\
human aging & 0.092 & 0.991 & high school mathematics & 0.380 & 0.886 \\
astronomy & 0.092 & 0.991 & high school physics & 0.426 & 0.906 \\
us foreign policy & 0.096 & 0.985 & abstract algebra & 0.535 & 0.847 \\
conceptual physics & 0.100 & 0.990 & moral scenarios & 0.562 & 0.901 \\
management & 0.102 & 0.984 & global facts & 0.572 & 0.697 \\
high school chemistry & 0.104 & 0.984 & college mathematics & 0.749 & 0.442 \\
\bottomrule
\end{tabular}
\end{table}

The table indicates that the general intelligence trait is highly correlated with most subjects. For example, professional psychology, prehistory, clinical knowledge, nutrition, and moral disputes all have sine values below \(0.02\) and Spearman correlations essentially equal to one. More broadly, many subjects in psychology, history, biology, medicine, law, and the social sciences have rank correlations above \(0.98\). These results provide empirical support for cross-domain borrowing: much of the comparative signal is shared across subjects, so estimating each subject entirely separately would ignore substantial shared information.

At the same time, the strength of that shared structure is not uniform. A smaller group of subjects shows a substantial departure from the general direction, indicating that complete pooling would be too restrictive. The most extreme case is college mathematics, with \(\sin(\widehat{\boldsymbol{\theta}}_{t},\widehat{\boldsymbol{\theta}}_{G})=0.749\) and a Spearman rank correlation of \(0.442\). Other distinctive subjects include global facts \((0.572,\,0.697)\), abstract algebra \((0.535,\,0.847)\), high school physics \((0.426,\,0.906)\), high school mathematics \((0.380,\,0.886)\), and machine learning \((0.319,\,0.847)\), where the two numbers in the parentheses give the sine discrepancy and Spearman rank correlation, respectively. These domains retain meaningful subject-specific ordering information beyond the dominant common factor. The evidence therefore points to an intermediate empirical regime: cross-domain borrowing is broadly useful, but the data still support substantial subject-specific departures for a smaller set of domains.

\begin{figure}[!h]
    \centering
    \includegraphics[width=\textwidth]{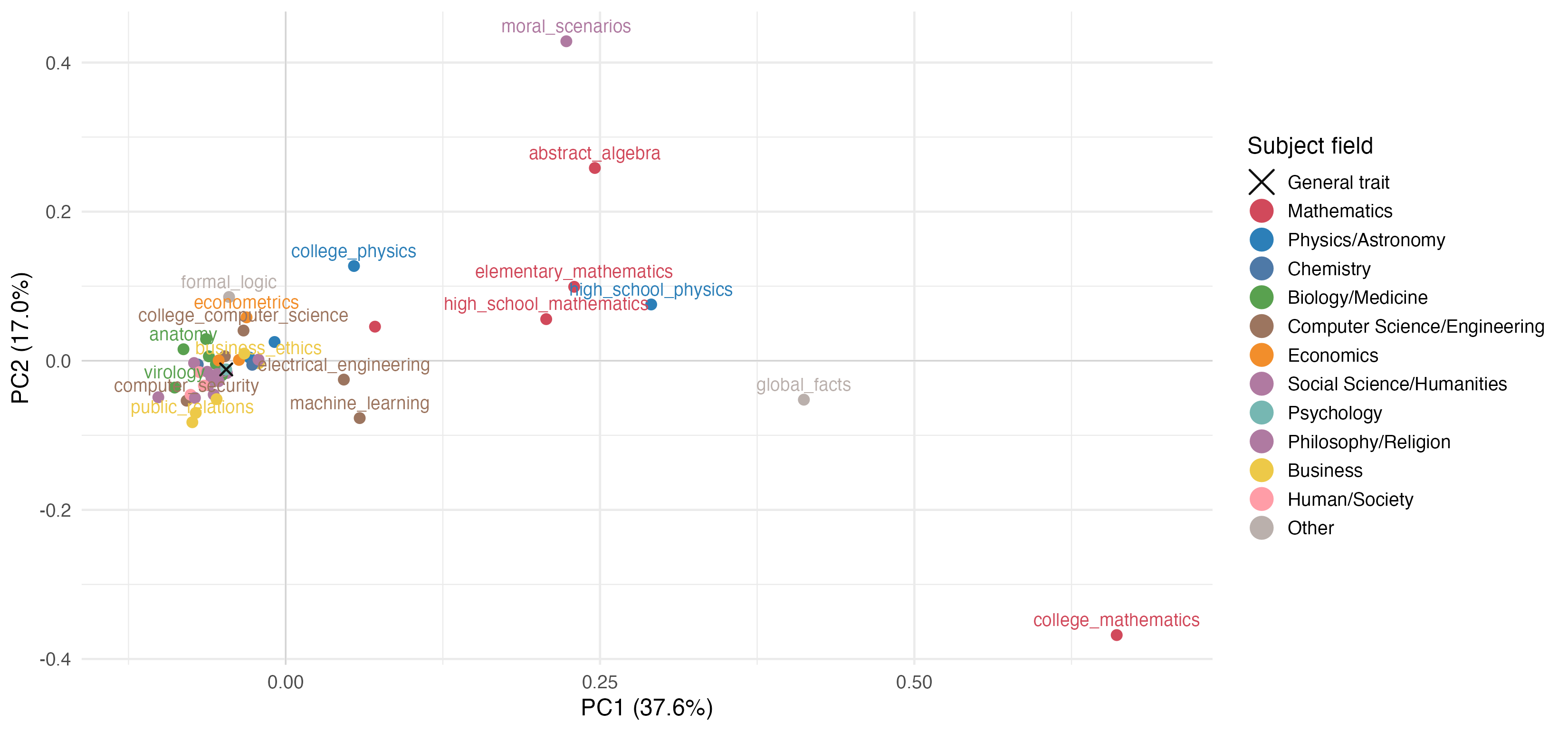}
    \caption{PCA of the normalized domain-specific trait vectors together with the normalized general intelligence trait. For clearer presentation, the 56 subjects are color-coded into broad field groups. To avoid excessive overlap and preserve readability, individual subject labels are omitted. In the axis labels, the percentages in parentheses report the proportion of total variance explained by PC1 and PC2, respectively.}
    \label{fig:domain-pca}
\end{figure}

Figure~\ref{fig:domain-pca} gives a complementary geometric summary of the between-subject heterogeneity by displaying the normalized trait vectors for all 56 subjects together with the normalized general intelligence trait in the same PCA projection. The first principal component of the normalized domain-specific trait vectors explains 37.6\% of the total variation, while the second explains 17.0\%. Most subjects remain in a relatively tight cluster around the general intelligence trait direction, consistent with the strong common component indicated by Table~\ref{tab:subject-rank-vs-general-rank}. A smaller set of subjects, however, separates systematically from this cluster. Along the first principal component, quantitatively intensive domains, including college mathematics, high school mathematics, high school physics, and college physics, show the clearest departures, suggesting that this direction primarily reflects quantitative specialization. The second principal component captures a different form of heterogeneity: abstract algebra and moral scenarios have relatively large positive coordinates, whereas college mathematics lies strongly in the opposite direction. Thus, subjects with similarly large discrepancies from the general intelligence trait can exhibit different patterns of specialization.

\subsection{General Intelligence and Overall LLM Ranking}

\begin{figure}[!h]
    \centering
      \includegraphics[width=\textwidth]{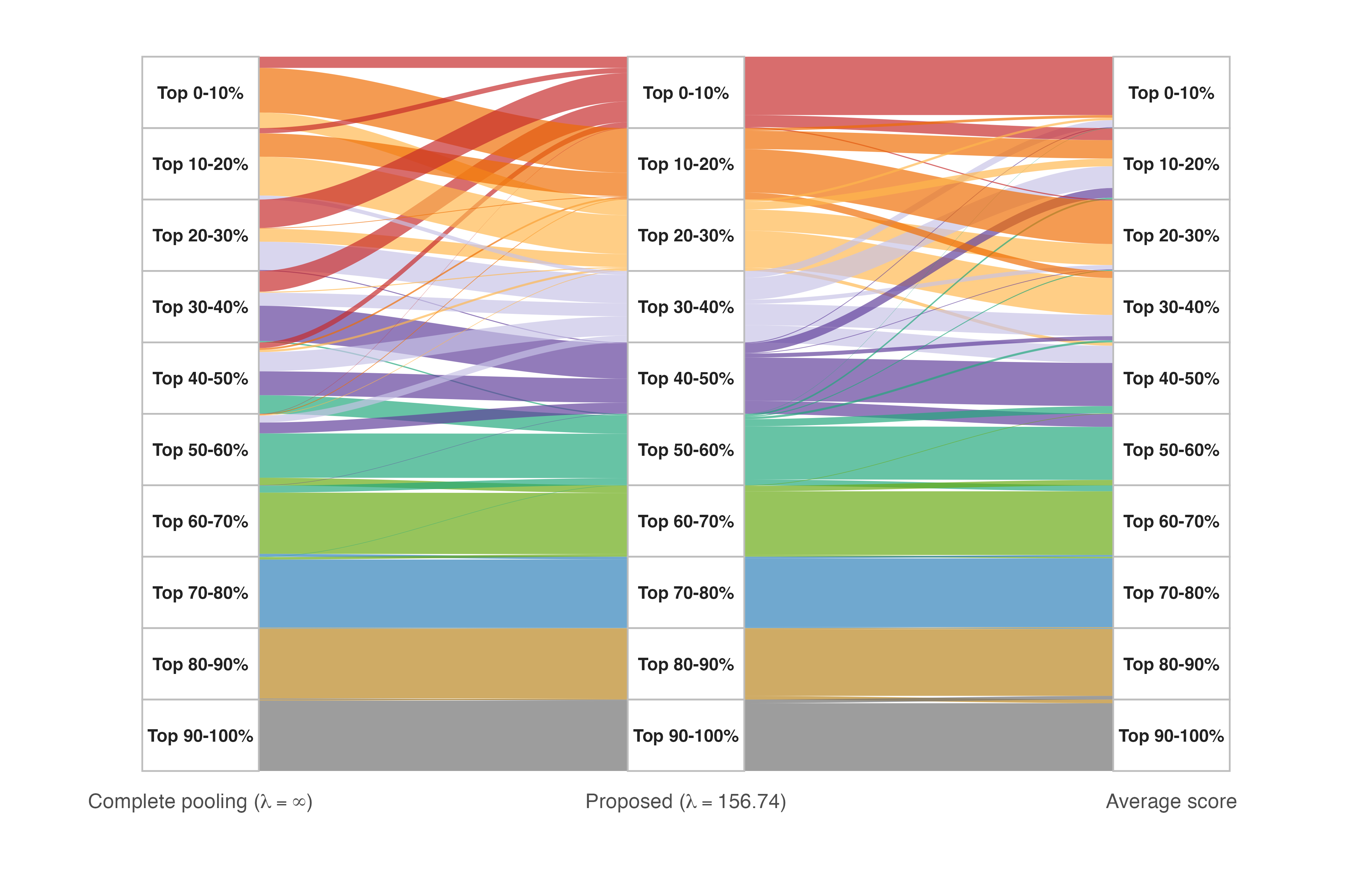}
    \caption{Tier-level ranking flow across three evaluation methods. The center column is the proposed estimator at the selected value \(\lambda=156.74\); the left column is the ranking based on complete pooling \((\lambda=\infty)\); and the right column is the direct average-score ranking. Colors are anchored to the proposed ranking tiers in the center column and then flow outward to the two comparison rankings.}
    \label{fig:ranking-flow}
\end{figure}

We next turn to the second research question and provide an overall ranking of the 4,272 models. We compare three rankings. The proposed ranking is induced by the estimated general intelligence trait \(\widehat{\boldsymbol{\theta}}_{G}^{\lambda}\) at the selected value \(\lambda=156.74\). The first alternative is an average-score ranking, which follows the leaderboard scoring rule by averaging subject-level accuracies across the 56 retained subjects. The second alternative is the ranking obtained under complete pooling, that is, under the constraint \(\boldsymbol{\theta}_1=\cdots=\boldsymbol{\theta}_{56}=\boldsymbol{\theta}_G\) induced by $\lambda = \infty$. Figure~\ref{fig:ranking-flow} compares the three rankings through a tier-level ranking flow, where models are grouped into deciles under each method. Colors are anchored to the proposed ranking in the center column, so that the flows to the left and right show how models within each proposed ranking tier are redistributed under complete pooling and average scoring, respectively. These two comparisons serve different purposes. Compared with average scoring, the proposed estimator isolates the effect of item characteristics. Relative to complete pooling, it isolates the effect of cross-domain heterogeneity in model capabilities.

We first compare the proposed ranking with the average-score ranking. On the one hand, the two rankings are largely consistent. The rankings have a Spearman rank correlation of 0.958 and share 19 of their top 20 models. On the other hand, we also see noticeable differences within the 10--50\% tiers, as shown in Figure~\ref{fig:ranking-flow}. Such differences may be attributable to the 
item-level measurement model. Unlike average scoring, which treats all item responses symmetrically, the proposed estimator incorporates item-level heterogeneity through the likelihood. In particular, the contribution of an item to latent-trait estimation depends on its estimated discrimination parameter $a_{tj}$, so that responses to more discriminating items carry greater information about differences in LLM capability. In our final estimate, the $a_{tj}$ parameters of 2,589 items (\(19.6\%\)) have been estimated to zero. For these items, the fitted response probability is independent of the trait and hence provides zero trait information. 
A further investigation of the items suggests that these items are substantially harder than the remaining items (17.1\% versus 64.5\% in mean accuracy), have a small negative mean correlation with the general intelligence trait (\(-0.189\) versus \(0.469\)), and are especially concentrated in virology, college chemistry, abstract algebra, formal logic, high school mathematics, and professional law. Some models that benefit in the average-score ranking from unusual success on these items move downward under the proposed estimator, whereas models with more coherent performance on the more positively discriminating items are moved upward.

We then compare the proposed ranking with complete pooling, i.e., the estimator when $\lambda = \infty$. The left side of Figure~\ref{fig:ranking-flow} shows that the two rankings are again strongly associated (Spearman correlation \(0.938\)), but the complete pooling estimator produces a much heavier downward tail. In particular, 271 LLMs fall by more than 1,000 positions, with their median rank changing from 234 under the proposed estimator to 1,399 under the complete-pooling estimator. The reason is not simply additional smoothing. As shown in the domain-specific analysis above, some subjects exhibit nontrivial departures from the common general direction. Once all domain-specific traits are forced to share the same ordering, items that discriminate positively under a subject-specific ordering may become non-monotone relative to the common ordering, and the pooled fit pushes many of their discrimination parameters to zero. Accordingly, about 30.0\% of the item discrimination parameters \(a_{tj}\) are estimated to be zero, compared with 19.6\% under the proposed fit. The LLMs with the largest downward shifts under the complete pooling estimator perform particularly well on these items, especially in subjects such as moral scenarios, professional law, and elementary mathematics. This comparison shows that allowing cross-domain heterogeneity is not a minor refinement: heterogeneity in subject-level comparative orderings is substantively present, and accounting for it materially changes the overall ranking.

Table~\ref{tab:top20-proposed} lists the top 20 LLMs under the proposed ranking together with their complete pooling and average-score ranks, and the full ranking is reported in Supplementary Materials C.4. In particular, the \texttt{AA05161*} series exhibits unusually strong performance, a point that was previously questioned in a Leaderboard discussion on Hugging Face \footnote{See
\href{https://huggingface.co/spaces/open-llm-leaderboard/open_llm_leaderboard/discussions/785}
{Open LLM Leaderboard Discussion No.~785}.}. Setting aside potential benchmark contamination, the proposed estimator provides a principled ranking by accounting for differences in item informativeness and cross-domain heterogeneity.

\begin{table}[!htbp]
\centering
\footnotesize
\caption{Top 20 LLMs under the proposed general intelligence trait and their ranks under the two comparison methods. Model type was determined from the Open LLM Leaderboard metadata and supplemented using information from the corresponding official model cards on the Hugging Face Hub when the metadata were missing or inconsistent.}
\label{tab:top20-proposed}
\resizebox{\textwidth}{!}{%
\begin{tabular}{clccccc}
\toprule
Proposed Rank & Model & Complete Pooling Rank & Average-Score Rank
& Architecture & Params (B) & Type \\
\midrule
1  & \texttt{AA051611/O0201}             & 33  & 1  & Llama & NA & Fine-tuned \\
2  & \texttt{AA051611/V0201}             & 197 & 2  & Llama & NA & Fine-tuned \\
3  & \texttt{AA051612/A0125}             & 529 & 4  & Llama & 34 & Fine-tuned \\
4  & \texttt{AA051615/A0204}             & 530 & 3  & Llama & 34 & Fine-tuned \\
5  & \texttt{AA051612/B0121}             & 435 & 7  & Llama & 34 & Fine-tuned \\
6  & \texttt{CausalLM/34b-beta}          & 727 & 6  & Llama & 34 & Fine-tuned \\
7  & \texttt{AA051611/V0202}             & 827 & 5  & Llama & 34 & Fine-tuned \\
8  & \texttt{AA051610/A0127}             & 688 & 8  & Llama & 34 & Fine-tuned \\
9  & \texttt{AA051615/A0304}             & 536 & 10 & Llama & 34 & Fine-tuned \\
10 & \texttt{AA051612/A0126}             & 668 & 11 & Llama & 34 & Fine-tuned \\
11 & \texttt{AA051610/O0128}             & 639 & 15 & Llama & NA & Fine-tuned \\
12 & \texttt{AA051615/A0221}             & 705 & 9  & Llama & 34 & Fine-tuned \\
13 & \texttt{AA051615/A0308-G}           & 685 & 12 & Llama & 34 & Fine-tuned \\
14 & \texttt{AA051612/A0124}             & 737 & 14 & Llama & 34 & Fine-tuned \\
15 & \texttt{NeverSleep/CausalLM-RP-34B} & 664 & 16 & Llama & 34 & Fine-tuned \\
16 & \texttt{AA051615/L0225}             & 670 & 17 & NA    & NA & Fine-tuned \\
17 & \texttt{Qwen/Qwen2-72B}             & 734 & 13 & Qwen2 & 72 & Pretrained \\
18 & \texttt{AA051615/L0223}             & 782 & 18 & NA    & NA & Fine-tuned \\
19 & \texttt{AA051612/B0122}             & 837 & 19 & Llama & 34 & Fine-tuned \\
20 & \texttt{AA051615/A0306}             & 854 & 22 & Llama & 34 & Fine-tuned \\
\bottomrule
\end{tabular}
}
\end{table}

\section{Discussion}\label{sec:discussion}

In this paper, we introduced a principled statistical framework for evaluating LLMs across multiple benchmark domains by integrating representation MTL with multidimensional IRT. Traditional leaderboard evaluation practices often rely on naive score aggregations or independent domain-specific estimates, which fail to account for heterogeneity in item counts and psychometric properties, as well as the strong cross-domain correlations inherent in modern LLM performance data. To address these limitations, our framework formulates an overall general intelligence trait alongside domain-specific competencies, incorporating a penalty regularizer to borrow information across domains. We established the theoretical optimality of the proposed estimators under suitable asymptotic regimes.  Simulations and an empirical application to the Hugging Face Open LLM Leaderboard demonstrated that our approach yields more reliable, robust, and accurate capability measures and rankings than competing baseline methods.

Despite the strong performance and theoretical guarantees of the current framework, several promising avenues remain open for future research. First, real-world benchmark datasets frequently contain flawed, ambiguous, or adversarially corrupted test items that misbehave and deviate from standard IRT assumptions (acting as outliers). Future work 
may develop robust estimation techniques, such as incorporating robust loss functions or sparse outlier detection mechanisms, to ensure that measurements remain reliable in the presence of anomalous items. Second, as the MMLU data analysis suggests, a single general intelligence trait may be insufficient to capture cross-domain heterogeneity. 
Therefore, it may be worth introducing additional structure into the current framework to account for the remaining unexplained variation, especially in the quantitative domains. 
One promising direction is to incorporate a hierarchical factor structure, reminiscent of traditional factor analysis, and determine its configuration in a data-driven manner \citep{Qiaoeal2025}.
Finally, the current framework evaluates LLMs on a fixed, predetermined set of items. Adapting this methodology to an active-learning setting, analogous to Computerized Adaptive Testing (CAT) in psychometrics \citep{wainer2000computerized}, would enable the evaluation process to dynamically select the most informative items for a given LLM. This extension would substantially reduce evaluation costs and token overhead while preserving high measurement precision.

\section*{Supplementary Materials}
The Supplementary Materials contain additional theoretical, simulation, and empirical results. 
Section A provides the proofs of the theoretical results, including the extension to the missing-data setting, auxiliary lemmas, sufficient conditions for Assumption~3, verification of these conditions under a random construction of the domain-specific traits, and the minimax lower bound. 
Section B provides additional details of the simulation study, including the cross-validation procedure for selecting the tuning parameter, verification of the theoretical conditions under the data-generation mechanism, and implementation details of the optimization algorithm and constraint enforcement. 
Section C provides supplementary results for the empirical application, including descriptive information on the LLMs in the leaderboard, selection of the tuning parameter, the relationship between the estimated intelligence traits and LLM size, and the full ranking of the LLMs based on the estimated general intelligence trait from the proposed method.

\section*{Acknowledgements}

This paper resulted from a grant funded by the
LSE-NYU Research Seed Fund 2025-2026.

\vspace{1.5cm}
\setstretch{1}
\bibliographystyle{apalike}  
\bibliography{ref}
\end{document}